\documentclass[11pt]{article}

\usepackage{geometry}                
\usepackage{graphicx}
\usepackage{amsmath,amssymb,amsfonts,amsthm,bm}
\usepackage{tikz-cd} 
\usepackage{cite}
\usepackage{authblk} 

\newcommand{\vol}{\bm{\mathrm{{vol}}}}

\newtheorem{prop}{Proposition}

\newtheorem{lemma}{Lemma}

\title{A Hamiltonian and geometric formulation of general Vlasov-Maxwell-type models}
\author[,1]{William Barham\thanks{Corresponding author: william.barham@utexas.edu}}
\author[,2]{Philip J. Morrison\thanks{Corresponding author: morrison@physics.utexas.edu}}
\author[3,4]{Eric Sonnendr{\"u}cker}
\affil[,1]{Oden Institute for Computational Engineering and Sciences, The University of Texas at Austin}
\affil[,2]{Department of Physics and Institute for Fusion Studies, The University of Texas at Austin}
\affil[3]{Max-Planck-Institut f{\"u}r Plasmaphysik}
\affil[4]{Technische Universit{\"a}t M{\"u}nchen, Zentrum Mathematik}
\date{\today}                     
\begin{document}

\maketitle

\section*{Abstract}

Three geometric formulations of the Hamiltonian structure of the macroscopic Maxwell equations are given: one in terms of the double de Rham complex, one in terms of $L^2$ duality, and one utilizing an abstract notion of duality. The final of these is used to express the geometric and Hamiltonian structure of kinetic theories in general media. The Poisson bracket so stated is explicitly metric free. Finally, as a special case, the Lorentz covariance of such kinetic theories is investigated. We obtain a Lorentz covariant kinetic theory coupled to nonlinear electrodynamics such as Born-Infeld or Euler-Heisenberg electrodynamics. 

\tableofcontents

\section{Introduction}
A host of electromagnetic phenomena occur in polarized and magnetized media. As the rationale of introducing polarization and magnetization amounts to the modeling of  complicated microscopic behavior in constitutive laws, the equations describing electromagnetism in a medium are often called the macroscopic Maxwell equations. Typically, an empirical linear model is used for these constitutive models. However, in many plasma models, it is useful to consider a self consistent model that can account for more complex couplings between the material, e.g. a charged particle model, and the fields. A systematic theory for lifting particle models to kinetic models and the Hamiltonian structure of these lifted models was given in \cite{morrison_gauge_free_lifting}. It has been shown that many kinetic models of interest fit into this framework such as guiding center drift kinetics \cite{morrison_gauge_free_lifting} and gyrokinetics \cite{BURBY20152073}, while a Lie-transform interpretation of the framework was given in \cite{Brizard-etal}.

We investigate the geometric structure of such kinetic models beginning with a detailed investigation of various ways of expressing the macroscopic Maxwell equations in a geometric language before considering the full kinetic theory.  Attention is paid to the geometric description of  orientation, which is accounted for in classical tensor  analysis (e.g.\ \cite{Synge-Schild}) by the introduction of pseudo-vectors, relative tensors,  tensor densities, etc., which in the language of split exterior calculus distinguishes between straight (orientation independent) and twisted (orientation dependent)  differential forms \cite{burke, frankel_2011}. In particular, \cite{hehl2012foundations} provides a cogent axiomatic derivation of classical electrodynamics which displays the significance of a geometric perspective. This language has recently been used in the context of geophysical fluids in \cite{eldred_and_bauer}.  Here, we first give a geometric statement of the Hamiltonian structure of Maxwell's equations in this  language of split exterior calculus. Next, we give a  second formulation that simplifies the model by expressing all duality structures in terms of the $L^2$ inner product. This yields a formulation frequently seen in finite element literature \cite{hiptmair_maxwell_cts_disc, ArnoldDouglasN2010Feec}. Finally, we present a formulation based on an abstract notion of duality. This final formulation has the advantage of expressing the structure of Maxwell's equations in a manner that clearly separates which structures depend on the metric tensor through the $L^2$ inner product (the Hamiltonian), and those structures which depend only on the natural pairing (the Poisson bracket). In addition to revealing the beautiful mathematics underpinning the Hamiltonian structure of these models, such fastidious attention to the duality structures at play in these models provides a solid foundation for their discretization by finite element methods which will be the subject of a future paper. 

Following this study of the geometric structure of the macroscopic Maxwell equations alone, we consider the full kinetic theory in general media of \cite{morrison_gauge_free_lifting}. We place this class of theories, including the Vlasov-Maxwell system, for the first time in a complete geometric framework. In particular, we derive a formulation which transparently demonstrates the metric free character of the Poisson bracket. Further discussion of kinetic theories in general media, including guiding center and gyrokinetic approximations, and a general methodology for discretizing such models may be found in \cite{doi:10.1063/1.4976849}. As a special case of the kinetic model in general media, we conclude by studying the Lorentz invariance of the Vlasov equation coupled to various models in nonlinear electrodynamics.

\section{A geometric formulation of Maxwell's equations}
\label{geoMEs}
A Hamiltonian formulation of the macroscopic Maxwell equations, as a component of a larger kinetic model, was given in \cite{morrison_gauge_free_lifting}. In this section, we consider the Maxwell component of this model in isolation and the various geometric interpretations one might give to the electromagnetic fields. The macroscopic Maxwell equations without free charge and current may be written
\begin{equation}
	\begin{split}
		\frac{\partial \bm{B}}{\partial t} &= - c \nabla \times \bm{E} \\
		\nabla \cdot \bm{B} &= 0
	\end{split}
	\quad
	\begin{split}
		\frac{\partial \bm{D}}{\partial t} &= c \nabla \times \bm{H} \\
		\nabla \cdot \bm{D} &= 0.
	\end{split}
\end{equation}
Hence, the $(\bm{D}, \bm{B})$ fields are the time-evolving fields whereas $(\bm{E}, \bm{H})$ may be related to the evolving fields through a general constitutive law:
\begin{equation}
	\bm{E} = \bm{E}(\bm{D}, \bm{B}) 
	\quad \text{and} \quad
	\bm{H} = \bm{H}(\bm{D}, \bm{B}).
\end{equation}

Define an energy functional
\begin{equation}
	K[ \bm{E}, \bm{B} ] = \int_Q \mathcal{K}(\bm{x}, \bm{E}, \bm{B}, \nabla \bm{E}, \nabla \bm{B}, ... )
	 \, \mathsf{d}^3 \bm{x}
\,,
\end{equation}
where $\mathsf{d}^3 \bm{x}$ is a  volume element of configurations space $Q$, 
so the macroscopic fields are given by
\begin{equation} \label{eq:macroscopic_fields}
	\bm{D} = \bm{E} - 4\pi \bm{P}(\bm{E}, \bm{B}) = \bm{E} - 4 \pi \frac{\delta K}{\delta \bm{E}} 
	\quad \text{and} \quad 
	\bm{H} = \bm{B} + 4\pi \bm{M}(\bm{E}, \bm{B}) = \bm{B} + 4 \pi \frac{\delta K}{\delta \bm{B}}\,.
\end{equation}
That is, we define the polarization and magnetization through the functional derivative of a general functional. This does not impede generality, and proves useful for specifying the Hamiltonian structure \cite{morrison_gauge_free_lifting}.

Further, let the Hamiltonian be
\begin{equation}
	H[\bm{E},\bm{B}] = K - \int_Q \bm{E} \cdot \frac{\delta K}{\delta \bm{E}}  \,\mathsf{d}^3 \bm{x}
  + \frac{1}{8 \pi} \int_Q ( \bm{E} \cdot \bm{E} + \bm{B} \cdot \bm{B} ) \,\mathsf{d}^3 \bm{x}
\,,
\end{equation}
and the Poisson bracket be
\begin{equation}
	\{ F, G \} = 4 \pi c \int_Q \left[ \frac{\delta F}{\delta \bm{D}} \cdot \nabla \times \frac{\delta G}{ \delta \bm{B}} 
				- \frac{\delta G}{\delta \bm{D}} \cdot \nabla \times \frac{\delta F}{\delta \bm{B}} \right]  \,\mathsf{d}^3 \bm{x}
\end{equation}
where all functional derivatives are understood in terms of the $L^2$ inner product. Then, as shown in \cite{morrison_gauge_free_lifting}, letting $\overline{H}[\bm{D}, \bm{B}] = H[\bm{E}, \bm{B}]$ and using the fact that (see \ref{appendix:deriv-of-ham} for the details)
\begin{equation}
	\frac{\delta \overline{H}}{\delta \bm{D}} = \frac{\bm{E}}{4 \pi} 
	\quad \text{and} \quad
	\frac{\delta \overline{H}}{\delta \bm{B}} = \frac{\bm{H}}{4 \pi},
\end{equation}
we recover the macroscopic Maxwell equations in Poisson bracket form: $\partial \bm{B}/\partial t=\{\bm{B},H\}$ and $\partial \bm{D}/\partial t=\{\bm{D},H\}$. The non-evolving Gauss constraints are Casimir invariants of the bracket, i.e., they represent quantities $C$ such that $\{C,F\}=0$ for all functionals $F$.

In Secs.~\ref{ssec:doubleDR}, \ref{ssec:duality}, and \ref{ssec:genpair} we will present three geometrical formulations of this model.  A central concern will be the metric dependence or independence of the Poisson bracket, that is, whether computation of the bracket between two general functionals requires knowing the metric tensor on configuration space. This is done in part to provide a foundation for future work on structure preserving discretizations of Maxwell's equations in general media and associated kinetic theories. A brief overview of the mathematical context and notational choices of this paper may be found in \ref{appendix:notation}. 

\subsection{Double de Rham complex formulation}
\label{ssec:doubleDR}

It is possible to directly translate the macroscopic Maxwell equations into the language of exterior calculus using the musical isomorphisms ($\flat,\sharp$) between vector fields and differential forms and the Hodge star operator ($\star$):
\begin{equation}
	\begin{split}
		\star \left(\frac{\partial \bm{B}}{\partial t}+ c \nabla \times \bm{E}  \right)^\flat &= 0 \\
		\star (\nabla \cdot \bm{B} ) &= 0
	\end{split}
	\hspace{4em}
	\begin{split}
		\star \left(\frac{\partial \bm{D}}{\partial t} - c \nabla \times \bm{H}  \right)^\flat &= 0 \\
		\star ( \nabla \cdot \bm{D}) &= 0\,,
	\end{split}
\end{equation}
which yields
\begin{equation} 
	\begin{split}
		\frac{\partial \bm{b}^2}{\partial t} &= -c \mathsf{d}_1 \bm{e}^1 \\
		\mathsf{d}_2 \bm{b}^2 &= 0
	\end{split}
	\hspace{4em}
	\begin{split}
		 \frac{\partial \tilde{\bm{d}}^2}{\partial t} &= c \tilde{\mathsf{d}}_1 \tilde{\bm{h}}^1 \\
		\tilde{\mathsf{d}}_2 \tilde{\bm{d}}^2 &= 0\,,
	\end{split}
\end{equation}
where we have identified the differential forms  $\bm{e}^1 = \bm{E}^\flat$, $\bm{b}^2 = \textbf{i}_{\bm{B}} \vol^3$, $\tilde{\bm{d}}^2 =  \textbf{i}_{\bm{D}} \vol^3$, and $\tilde{\bm{h}}^1 = \bm{H}^\flat$. Here we have replaced the volume element  $\mathsf{d}^3 \bm{x}
$ by the orientation respecting (twisted) volume form $ \vol^3$ (see \cite{frankel_2011}) with $\textbf{i}_{\bm{B}}$ being the interior product, and we have  used the appropriate exterior derivatives, e.g.,  $\mathsf{d}_1$.  (See \ref{appendix:notation} for further details.)  As $\bm{B}$ and $\bm{H}$ are pseudovectors (i.e.\ they change sign under orientation reversing coordinate transformations), $\bm{b}^2 = \textbf{i}_{\bm{B}} \vol^3$ is a straight 2-form while $\tilde{\bm{h}}^1 = \bm{H}^\flat$ is twisted 1-form, whence we use the tilde. Consistency requires that each equation only contain differential forms of like kind (straight or twisted) so that the form of the equations remain invariant under orientation reversing coordinate transformations. 

Adapting ideas from split exterior calculus (see \cite{eldred_and_bauer}), our first formulation of Maxwell's equations makes explicit use of the double de Rham complex and seeks to split the model into pieces which are metric dependent (the Hamiltonian) and pieces which are purely topological (the Poisson bracket). The Hamiltonian and Poisson bracket are written as follows:
\begin{equation} \label{geometric_hamiltonian}
	H[\bm{e}^1, \bm{b}^2] = K - \int_Q \frac{\delta K}{\delta \bm{e}^1} \wedge \star \bm{e}^1 + \frac{1}{8 \pi} \left[ \int_Q \bm{e}^1 \wedge \star \bm{e}^1 + \bm{b}^2 \wedge \star \bm{b}^2 \right]
\end{equation}
and
\begin{equation} \label{pb_diff_forms}
	\{ F, G \} = 4 \pi c \left[ \int_Q \left( \frac{\tilde{\delta} F}{\delta \tilde{\bm{d}}^2} \wedge \mathsf{d} \frac{\tilde{\delta} G}{\delta \bm{b}^2} 
			- \frac{\tilde{\delta} G}{\delta \tilde{\bm{d}}^2} \wedge \mathsf{d} \frac{\tilde{\delta} F}{ \delta \bm{b}^2 } \right) \right]\,,
\end{equation}
where the tildes indicate twisted functional derivatives, see \ref{appendix:notation}. Because the pushforward distributes over the wedge product, this bracket is explicitly metric free. On the other hand, the Hodge star operator contains metric information making the Hamiltonian metric dependent. 

Using methods very similar to those given in \ref{appendix:deriv-of-ham}, one may show that if we let $\overline{H}[\tilde{\bm{d}}^2, \bm{b}^2] = H[\bm{e}^1, \bm{b}^2]$, then
\begin{equation}
	D \overline{H}[ \tilde{\bm{d}}^2, \bm{b}^2](\delta \tilde{\bm{d}}^2, \delta \bm{b}^2) 
	= \int_Q \left(\frac{\bm{e}^1}{4 \pi} \wedge \delta \tilde{\bm{d}}^2 
		+ \frac{\tilde{\bm{h}}^1}{4 \pi} \wedge \delta \bm{b}^2 \right).
\end{equation}
Hence, it immediately follows that we recover the above equations of motion in Poisson bracket form and that the Gauss constraints are Casimirs of the bracket since $\mathsf{d}^2 = 0$. 

This formulation is attractive because of its partition of metric dependence and independence between the Hamiltonian and the bracket and because of its use of elementary objects from differential geometry. Moreover, the geometric significance of each variety of differential form (twisted and straight) can enhance  physical intuition (see \cite{burke, hehl2012foundations, e_tonti_1975}) and aids in the design of numerical methods \cite{bossavit_1988a, hiptmair_maxwell_cts_disc}. Numerical methods based on this modeling perspective typically explicitly discretize the Hodge star operator yielding a matrix that is, in general, neither symmetric positive definite nor even square \cite{hiptmair_discrete_hodge_star}; an exception to this usual shortcoming of discrete Hodge star operators may be found in \cite{kapidani_2022}. This is inconvenient as the discrete Hodge star operator should act as an inner product at the discrete level. Hence, it is often more convienient to base discrete duality structures entirely on the $L^2$ inner product \cite{bochev_and_hyman}. From a modeling perspective, the introduction of two distinct de Rham complexes which are in duality with each other via the Hodge star operator, while geometrically intuitive, is unnecessarily complicated. One may instead construct a formulation based on only a single duality structure (rather than the two needed to construct the Hodge star operator). 

\subsection{Formulation using $L^2$ duality}
\label{ssec:duality}

The model may be expressed entirely in terms of the $L^2$ inner product: 
\begin{equation}
	(\cdot, \cdot): \Lambda^k \times \Lambda^k \to \mathbb{R}
	\quad \text{where} \quad
	( \omega^k, \eta^k) = \int_Q \omega^k \wedge \star \eta^k.
\end{equation}
In this section, all functional derivatives will be understood to be identified with respect to this duality pairing. The Hamiltonian is written
\begin{equation}
	H[\bm{e}^1, \bm{b}^2] = K - \left( \frac{\delta K}{\delta \bm{e}^1}, \bm{e}^1 \right) + \frac{1}{8 \pi} \left[ \left( \bm{e}^1, \bm{e}^1 \right) + \left( \bm{b}^2, \bm{b}^2 \right) \right]\,, 
\end{equation}
while the Poisson bracket is written
\begin{equation}
	\{ F, G \} = 4 \pi c \left[ \left( \frac{\delta G}{\delta \bm{b}^2}, \mathsf{d}_1 \frac{\delta F}{\delta \bm{d}^1} \right) - \left( \frac{\delta F}{ \delta \bm{b}^2 }, \mathsf{d}_1 \frac{\delta G}{\delta \bm{d}^1} \right) \right]
\end{equation}
and the constitutive laws are given by
\begin{equation}
	\bm{d}^1 = \bm{e}^1 - 4 \pi \frac{\delta K}{\delta \bm{e}^1} \quad \text{and} \quad \bm{h}^2 = \bm{b}^2 + 4 \pi \frac{\delta K}{\delta \bm{b}^2}.
\end{equation}
Again, following an approach nearly identical to that given in \ref{appendix:deriv-of-ham}, one may show that
\begin{equation}
	\frac{\delta \overline{H}}{\delta \bm{d}^1} = \frac{\bm{e}^1}{4 \pi} 
	\quad \text{and} \quad 
	\frac{\delta \overline{H}}{\delta \bm{b}^2} = \frac{ \bm{b}^2}{4 \pi} + \frac{\delta K}{\delta \bm{b}^2} = \frac{\bm{h}^2}{4 \pi}.
\end{equation}
Hence, it follows that for any functional $F = F[\bm{d}^1, \bm{b}^2]$,
\begin{equation}
	\begin{split}
		\dot{F} 
			= \{ F, \overline{H} \} 
			= c \left[ \left( \frac{ \delta F}{\delta \bm{d}^1}, \mathsf{d}^* \bm{h}^2 \right) 
				- \left( \mathsf{d} \bm{e}^1, \frac{\delta F}{\delta \bm{b}^2} \right) \right] 
	\end{split}
	\quad \implies \quad
	\begin{split}
		\partial_t \bm{d}^1 &= c \mathsf{d}^* \bm{h}^2 \\
		\partial_t \bm{b}^2 &= - c \mathsf{d} \bm{e}^1\,,
	\end{split}
\end{equation}
where $(\omega, \mathsf{d} \eta) = (\mathsf{d}^* \omega, \eta)$. 

This formulation is somewhat simpler than the previous of Sec.~\ref{ssec:doubleDR} and utilizes only one duality structure on the differential forms (the $L^2$ inner product). However, the Poisson bracket so expressed is deficient in that, so expressed, it changes form under coordinate transformation due to the metric dependence of the $L^2$ inner product. The Poisson bracket is a purely topological quantity, and therefore should possess a metric free expression. As we shall see in the next section, this apparent dependence of the Poisson bracket on the metric cancels out if we identify functional derivatives with the natural duality pairing. 

\subsection{Formulation with abstract duality pairing}
\label{ssec:genpair}

This final formulation avoids explicitly identifying the dual space using a duality structure instead leaving duality abstract and general. This yields a model that is more descriptive and general than the previous two, but also requires more care regarding the functional analytic context. 

Let $(X, (\cdot, \cdot)_X)$ be a Hilbert space and let $X^* \sim X$ denote its dual space. Moreover, being Hilbert, the space is reflexive so that $X^{**} \sim X$. Let $f \colon X \to \mathbb{R}$. We denote the Fr{\'e}chet derivative at $v \in X$ in the direction $u \in X$ by 
\begin{equation*}
	D_X f [v] u = \left\langle \frac{\delta f}{\delta v}, u \right\rangle_{X^*, X}.
\end{equation*}
By the Riesz representation theorem, there exists a linear isomorphism $\mathcal{R}\colon X \to X^*$ such that
\begin{equation*}
	\langle \mathcal{R} u, v \rangle_{X^*, X} = (u, v)_X.
\end{equation*}
Let $f_* \colon X^* \to \mathbb{R}$ and define
\begin{equation*}
	D_{X^*} f_* [v_*] u_* = \left\langle u_*, \frac{\delta f_*}{\delta v_*} \right\rangle_{X^*, X},
\end{equation*}
where we have used reflexivity of $X$. Let $\mathcal{R}u = u_*$, $\mathcal{R}v = v_*$, and $f = f_* \circ \mathcal{R}$ so that
\begin{equation}
	f[v] = f_*[v_*] 
	\quad \text{and} \quad
	D_X f [v] u = D_{X^*} f_* [v_*] u_*\,.
\end{equation}
From this, we find that
\begin{equation}
	\begin{aligned}
		\left\langle u_*, \frac{\delta f_*}{\delta v_*} \right\rangle_{X^*, X} &=  \left\langle \frac{\delta f}{\delta v}, u \right\rangle_{X^*, X} 
			\implies \left( \mathcal{R}^{-1} u_*, \frac{\delta f_*}{\delta v_*} \right)_X = \left( \mathcal{R}^{-1} \frac{\delta f}{\delta v}, u \right)_X \\
		&\implies \left( u, \frac{\delta f_*}{\delta v_*} \right)_X = \left( u, \mathcal{R}^{-1} \frac{\delta f}{\delta v} \right)_X 
			\implies  \mathcal{R}^{-1} \frac{\delta f}{\delta v} = \frac{\delta f_*}{\delta v_*}.
	\end{aligned}
\end{equation}
Hence, one may translate expressions involving functional derivatives with respect to variables on the primal space to functional derivatives with respect to variables on the dual space using the Riesz map. We shall use this to write the macroscopic Maxwell Poisson bracket abstractly without reference to a metric. 

The spaces of differential $k$-forms must be Hilbert in order for our theory to be consistent. Hence, we specify that
\begin{equation}
	V^k = H^1 \Lambda^k(\Omega) := \{ \omega \in L^2 \Lambda^k(\Omega) : \mathsf{d}_k \omega \in L^2 \Lambda^{k+1}(\Omega) \}.
\end{equation}
Let $\mathcal{R}_k: V^k \to (V^k)^*$ denote the Riesz map on $k$-forms. We define
\begin{equation}
	\bm{d}^1_* = \mathcal{R}_1 \bm{e}^1 - 4 \pi \frac{\delta K}{\delta \bm{e}^1} \in (V^1)^* \quad \text{and} \quad \bm{h}^2_* = \mathcal{R}_2 \bm{b}^2 - 4 \pi \frac{\delta K}{\delta \bm{b}^2} \in (V^2)^*
\end{equation}
where the functional derivatives are identified with the natural pairing so that they live in $(V^k)^*$, and we define
\begin{equation}
	\bm{d}_*^1 = \mathcal{R}_1 \bm{d}^1 \quad \text{and} \quad \bm{h}_*^2 = \mathcal{R}_2 \bm{h}^2.
\end{equation}
Then we may write the Hamiltonian as
\begin{equation}
	H[\bm{e}^1, \bm{b}^2] = K - \left\langle \frac{\delta K}{\delta \bm{e}^1}, \bm{e}^1 \right\rangle_{(V^1)^*, V^1} + \frac{1}{8 \pi} \left[ \left( \bm{e}^1, \bm{e}^1 \right)_{L^2} + \left( \bm{b}^2, \bm{b}^2 \right)_{L^2} \right]
\end{equation}
and the Poisson bracket may be written
\begin{equation}
	\{ F, G \} = 4 \pi c \left[ \left\langle \frac{\delta G}{\delta \bm{b}^2}, \mathsf{d}_1 \frac{\delta F}{\delta \bm{d}^1_*} \right\rangle_{(V^2)^*, V^2}
					- \left\langle \frac{\delta F}{ \delta \bm{b}^2 }, \mathsf{d}_1 \frac{\delta G}{\delta \bm{d}^1_*} \right\rangle_{(V^2)^*, V^2} \right]\,,
\end{equation}
where we have made use of the fact that $\delta F/\delta \bm{d}^1_* \in (V^1)^{**} \sim V^1$. This Poisson bracket is metric-free because duality is expressed through functional evaluation which is coordinate independent. 

Therefore, letting $\overline{H}[\bm{d}^1_*, \bm{b}^2] = H[\bm{e}^1, \bm{b}^2]$ and using the chain rule, we find that
\begin{equation}
	\mathcal{R}_1^{-1} \left( \frac{\delta \overline{H}}{\delta (\mathcal{R}_1^{-1} \bm{d}^1_*)} \right) = \frac{\bm{e}^1}{4 \pi} 
	\quad \text{and} \quad 
	\mathcal{R}_2^{-1} \left( \frac{\delta \overline{H}}{\delta \bm{b}^2} \right) = \frac{ \bm{b}^2}{4 \pi} + \mathcal{R}_2^{-1} \left( \frac{\delta K}{\delta \bm{b}^2} \right) = \frac{\mathcal{R}_2^{-1}(\bm{h}^2)}{4 \pi}\,.
\end{equation}
Letting $\overline{H}_*[\bm{d}^1_*, \bm{b}^2] = \overline{H}[\bm{d}^1, \bm{b}^2] = H[\bm{e}^1, \bm{b}^2]$, we find that
\begin{equation}
	\frac{\delta \overline{H}_*}{\delta \bm{d}^1_*} = \frac{\bm{e}^1}{4 \pi} 
	\quad \text{and} \quad 
	\frac{\delta \overline{H}_*}{\delta \bm{b}^2} = \frac{1}{4 \pi} \left( \mathcal{R}_2 \bm{b}^2 + \frac{\delta K}{\delta \bm{b}^2} \right) = \frac{\bm{h}^2_*}{4 \pi}\,.
\end{equation}
Hence, for arbitrary functionals $F$ of the observables, we find
\begin{equation}
	\dot{F} = \{F, \overline{H} \} = c \left[ \left\langle \bm{h}^2_*, \mathsf{d}_1 \frac{\delta F}{\delta \bm{d}^1_*} \right\rangle - \left\langle \frac{\delta F}{\delta \bm{b}^2}, \mathsf{d}_1 \bm{e}^1 \right\rangle \right].
\end{equation}
Thus, we have a metric free representation of the Poisson bracket by identifying the variables $(\bm{d}^1_*, \bm{h}^2_*)$ with the dual space. This is entirely natural as the constitutive relations are prescribed by functional derivatives which themselves naturally live in the dual space. 

As a final note, the Riesz map is specified by the natural inner product on the Hilbert space. However, because
\begin{equation}
	H^1 \Lambda^k(\Omega) \subset L^2 \Lambda^k(\Omega) 
	\implies 
	L^2 \Lambda^k(\Omega) \sim (L^2 \Lambda^k(\Omega))^* \subsetneq (H^1 \Lambda^k(\Omega))^*,
\end{equation}
it follows that sufficiently regular functionals $K$ might have their functional derivatives identified with the primal space through the $L^2$ pairing rather than the natural Riesz map. This reduces the theory to one which is equivalent to the previous formulation in terms of $L^2$ duality. Thus, this formulation may be seen as a generalization of the $L^2$ theory that accommodates polarizations and magnetizations which cannot be identified as an element of $L^2 \Lambda^k(\Omega)$. Because of its generality and its partition of the metric dependent and independent components of the theory, this modeling paradigm provides a convenient starting place for a finite element discretization of the macroscopic Maxwell equations in Hamiltonian form. This will be the subject of future work.

\subsection{Some polarization examples}
We briefly consider some of the kinds of models that might be furnished by this modeling framework. First, consider an intensity dependent index of refraction:
\begin{equation}
	\bm{P} = \left( \chi^1 + \chi^3 | \bm{E} |^2 \right) \bm{E} \iff \bm{p}^1 = \left( \chi^1 + \chi^3 | \bm{e}^1 |^2 \right) \bm{e}^1\,,
\end{equation}
where $\chi^1$ and $\chi^3$ are scalars for simplicity. Such a model accounts for the lowest order nonlinear effects found in noncentrosymmetric media, and has been used to account for laser self-focusing in plasmas (see e.g.\  \cite{BoydNLO} and \cite{ShenNLO}). The $K$ functional leading to this polarization is
\begin{equation}
	K = - \int_Q \left( \frac{\chi^1}{2} | \bm{E} |^2 +  \frac{\chi^3}{4} | \bm{E} |^4 \right) \mathsf{d}^3 \bm{x}
 \iff K = - \int_Q \left( \frac{\chi^1}{2} | \bm{e}^1 |^2 + \frac{\chi^3}{4} | \bm{e}^1 |^4 \right) \mathsf{d}^3 \bm{x}
.
\end{equation}
We find that the Hamiltonian of such a system is given by
\begin{equation}
	\begin{aligned}
		H &= \frac{1}{8 \pi} \int_Q \Big[ \left( 1 + 4 \pi \chi^1 \right) | \bm{E} |^2 + | \bm{B} |^2 + 6 \pi \chi^3 | \bm{E} |^4 \Big] \mathsf{d}^3 \bm{x}
 \\
			&= \frac{1}{8 \pi} \int_Q \Big[ \left( 1 + 4 \pi \chi^1 \right) | \bm{e}^1 |^2 + | \bm{b}^2 |^2 + 6 \pi \chi^3 | \bm{e}^1 |^4 \Big] \mathsf{d}^3 \bm{x}\,.
	\end{aligned}
\end{equation}
We could proceed in an analogous manner for the magnetic field.

As a second example, we might consider a system where the polarization depends on the electric field nonlocally in space. For example,
\begin{equation}
	\bm{P} = \alpha \bm{E} + \beta \Delta \bm{E} 
	\iff
	\bm{p}^1 = \alpha \bm{e}^1 + \beta \left( \mathsf{d} \mathsf{d}^* + \mathsf{d}^* \mathsf{d} \right) \bm{e}^1.
\end{equation}
For example, in one dimension  such a polarization might arise from an energy functional with a nonlocal kernel:
\begin{equation}
	K = \iint_Q \chi(x - x') e^0(x)  e^0(x') \mathsf{d} x \mathsf{d} x' 
	\approx \alpha (e^0, e^0) + \beta ( \mathsf{d}_0 e^0, \mathsf{d}_0 e^0 )
\end{equation}
where $\hat{\chi}(0) = \alpha$, $\hat{\chi}'(0) = 0$, and $\hat{\chi}''(0) = \beta$ (the hat indicates the Fourier transform). The energy functional for such nonlocal polarizations may be written
\begin{equation}
	\begin{aligned}
		K &= - \frac{1}{2} \int_Q \left[ \alpha | \bm{E} |^2 + \beta \left( | \nabla \cdot \bm{E} |^2 + | \nabla \times \bm{E} |^2 \right) \right] \mathsf{d}^3 \bm{x} \\
			&= - \frac{1}{2} \int_Q \left[ \alpha | \bm{e}^1 |^2 + \beta \left( | \mathsf{d}^* \bm{e}^1 |^2 + | \mathsf{d} \bm{e}^1 |^2 \right) \right] \mathsf{d}^3 \bm{x}
	\end{aligned}
\end{equation}
Hence, assuming homogeneous boundary conditions, one obtains the Hamiltonian
\begin{equation}
	\begin{aligned}
		H &= \frac{1}{8 \pi} \int_Q \Big[ \left(1 + 4 \pi \alpha \right) | \bm{E} |^2 + | \bm{B} |^2
			+ 4 \pi \beta \left( | \nabla \cdot \bm{E} |^2 + | \nabla \times \bm{E} |^2 \right) \Big] \mathsf{d}^3 \bm{x} \\
		&= \frac{1}{8 \pi} \int_Q \Big[ \left(1 + 4 \pi \alpha \right) | \bm{e}^1 |^2 + | \bm{b}^2 |^2
			+ 4 \pi \beta \left( | \mathsf{d}^* \bm{e}^1 |^2 + | \mathsf{d} \bm{e}^1 |^2 \right) \Big] \mathsf{d}^3 \bm{x}\,.
	\end{aligned}
\end{equation}
This yields a Maxwell wave equation of the form
\begin{equation}
	\partial_t^2 \left[ ( \alpha + \beta \Delta )^{-1} \bm{E} \right] + c^2 \nabla \times \nabla \times \bm{E} = 0.
\end{equation}
Restricting our attention temporarily to one-dimensional plane wave solutions of Maxwell's equations, we obtain the dispersion relation
\begin{equation}
	\omega(k) = \pm \sqrt{ \frac{ c^2 k^2}{ \alpha + \beta k^2 }} \implies v_g(k) = \pm \frac{\alpha}{c^2} \left( \frac{c^2}{\alpha + \beta k^2} \right)^{3/2}, \ v_{ph}(k) = \pm \sqrt{ \frac{c^2}{\alpha + \beta k^2}}.
\end{equation}
Hence, this Hamiltonian models a dispersive medium which retards the propagation of high wavenumber modes. We could proceed in an analogous manner to define a magnetization with nonlocal dependence on the magnetic field.

\section{A geometric Vlasov-Maxwell model in general media}
We now turn our attention to the geometric interpretation of the full kinetic model given in \cite{morrison_gauge_free_lifting} which extends the previously described model for Maxwell's equations with general, self-consistent polarization and magnetization to be coupled to a kinetic theory. We briefly review this model stated in the language of vector calculus (in Gaussian units) before proceeding. As before, we define an energy functional $K$ which acts as a coupling between the fields and the matter thus giving rise to the polarization and magnetization:
\begin{equation}
	K = K[f, \bm{E}, \bm{B}] = \int \mathcal{K}(\bm{x}, \bm{v}, \bm{E}, \bm{B}, \nabla_{\bm{x}} \bm{E}, \nabla_{\bm{x}} \bm{B}, \cdots) f(\bm{x}, \bm{v}) \mathsf{d}^3 \bm{x} \mathsf{d}^3 \bm{v}.
\end{equation} 
The $(\bm{D}, \bm{H})$ fields are defined as before in equation (\ref{eq:macroscopic_fields}). For notational convenience, the standard (finite dimensional) Poisson bracket and Littlejohn's bracket \cite{doi:10.1063/1.524053} are respectively denoted:
\begin{equation}
	[g,h]_{\bm{v}} = \frac{1}{m} \left( \nabla_{\bm{x}} g \cdot \nabla_{\bm{v}} h - \nabla_{\bm{x}} h \cdot \nabla_{\bm{v}} g \right)
	\quad \text{and} \quad
	[g,h]_{\bm{B}} = \frac{q}{m^2 c} \bm{B} \cdot \left( \nabla_{\bm{v}} g \times \nabla_{\bm{v}} h \right).
\end{equation}
Then the kinetic model may be written:
\begin{equation}
	\begin{aligned}
		&\partial_t f + [f, \mathcal{K}]_{\bm{v}} + [f, \mathcal{K}]_{\bm{B}} + \frac{q}{m} \bm{E} \cdot \nabla_{\bm{v}} f = 0 \\
		&\partial_t \bm{D} - c \nabla \times \bm{H} + \frac{4 \pi q}{m} \int \nabla_{\bm{v}} \mathcal{K} f \mathsf{d}^3 \bm{v} = 0 \\
		&\partial_t \bm{B} + c \nabla \times \bm{E} = 0.
	\end{aligned}
\end{equation}
This model possesses a Hamiltonian formulation. The Hamiltonian is given by
\begin{equation}
	H[f, \bm{E}, \bm{B}] = K - \int \frac{\delta K}{\delta \bm{E}} \cdot \bm{E} \mathsf{d}^3 \bm{x} + \frac{1}{8 \pi} \int \left( | \bm{E} |^2 + | \bm{B} |^2 \right) \mathsf{d}^3 \bm{x}.
\end{equation}
As was the case for the macroscopic formulation of Maxwell's equations, while the Hamiltonian is most naturally stated in terms of the fields $(\bm{E}, \bm{B})$, the Poisson bracket is most naturally stated in terms of $(\bm{D}, \bm{B})$:
\begin{equation}
	\begin{aligned}
		\{F, G\} 
		&= \int f \left[ \frac{\delta F}{\delta f}, \frac{\delta G}{\delta f} \right]_{\bm{v}} \mathsf{d}^3 \bm{x} \mathsf{d}^3 \bm{v}
			+ \int f \left[ \frac{\delta F}{\delta f}, \frac{\delta G}{\delta f} \right]_{\bm{B}} \mathsf{d}^3 \bm{x} \mathsf{d}^3 \bm{v} \\
		&\quad+ \frac{4 \pi q}{m} \int f \left( \frac{\delta G}{\delta \bm{D}} \cdot \nabla_{\bm{v}} \frac{\delta F}{\delta f} -
			\frac{\delta F}{\delta \bm{D}} \cdot \nabla_{\bm{v}} \frac{\delta G}{\delta f} \right) 
			\mathsf{d}^3 \bm{x} \mathsf{d}^3 \bm{v} \\
		&\quad+ 4 \pi c \int \left( \frac{\delta F}{\delta \bm{D}} \cdot \nabla \times \frac{\delta G}{\delta \bm{B}} 
			- \frac{\delta G}{\delta \bm{D}} \cdot \nabla \times \frac{\delta F}{\delta \bm{B}} \right) \mathsf{d}^3 \bm{x}
	\end{aligned}
\end{equation}
where $m$ and $q$ are the mass and charge of the plasma species in question. All functional derivatives in this formulation are identified with the $L^2$ inner product. It is helpful to establish some terminology. The Poisson bracket splits into four parts: the first is the Poisson bracket which gives rise to the Vlasov equation and is thus called the Vlasov bracket; the last, as we saw in the previous section, is the bracket for Maxwell's equations and is called the Maxwell bracket; the middle two are called the particle coupling brackets because they mediate the coupling between the fields and the plasma. 

\subsection{Translating the Hamiltonian structure into a geometric language}
We begin our investigation of this model's geometric formulation with a brief review of the metric free construction of the canonical Poisson bracket. This construction is classical (emerging from \cite{RevModPhys.36.572, mackey_quantum_mechanics}; see, e.g., \cite{bishop-goldberg}), so we refer the reader to these references for a detailed discussion and simply recall the definitions. If $Q$ is a manifold, let $\pi_Q: T^*Q \to Q$ be the cotangent bundle projection and $\pi_{T^*Q}: T(T^*Q) \to T^*Q$ be the tangent bundle projection of $T(T^*Q)$. Letting $a \in T(T^*Q)$, we define the canonical $1$-form, $\theta$, by the formula $\langle a, \theta \rangle = \langle T \pi_Q a, \pi_{T^*Q} a \rangle$ where $T\pi_Q: T(T^*Q) \to TQ$ is the tangent map of $\pi_Q$. The canonical symplectic $2$-form is then defined to be $\omega = -\mathsf{d} \theta$ and may be shown to be full rank. Finally, one defines the Poisson bivector as the inverse of $\omega$. That is, we define
\begin{equation}
	[f, g] = J(\mathsf{d} f, \mathsf{d} g) = \omega(X_f, X_g)
\end{equation}
where we define $\mathsf{i}_{X_f} \omega = \omega(X_f, \cdot) = \mathsf{d} f$ and similarly for $X_g$. Because of the non-degeneracy of $\omega$, this expression is well defined. Moreover, its construction made no use of a metric. This is clear when we write the canonical Poisson bracket in local coordinates $(\bm{x}, \bm{u}) \in T^*Q$: 
\begin{equation}
	\left[ f, g \right] = \frac{\partial f}{\partial x^i} \frac{\partial g}{\partial u_i} - \frac{\partial g}{\partial x^i} \frac{\partial f}{\partial u_i}.
\end{equation}

We call $Q$ the configuration space and let $\mathfrak{g} = C^\infty(T^*Q)$. Then we interpret the phase space density as living in $\mathfrak{g}^*$, i.e. $f \in \mathfrak{g}^*$. We might think of $\mathfrak{g}$ as the space of $0$-forms over $T^*Q$ and $\mathfrak{g}^*$ as the space of $6$-forms, however, we find that it is cleaner and more general to keep all notions of duality pairing abstract rather than commit to a single perspective of duality. As in \cite{MARSDEN1982394} (see also \cite{pjm98}), one may write the Lie-Poisson particle bracket as
\begin{equation}
	\{F, G\}_{LP} = \left\langle f, \left[ \frac{\delta F}{\delta f}, \frac{\delta G}{\delta f} \right] \right\rangle_{\mathfrak{g}^*, \mathfrak{g}}\,, 
\end{equation}
where $[\cdot, \cdot]$ is the canonical Poisson bracket and functional derivatives with respect to $f$ are understood in terms of the natural duality pairing (functional evaluation) between $\mathfrak{g}$ and $\mathfrak{g}^*$. Therefore, because $f \in \mathfrak{g}^*$, it follows that $\delta F/\delta f \in \mathfrak{g} = C^\infty(T^*Q)$. 

Recall from section \ref{ssec:genpair} that $\bm{d}^1_* \in (\Lambda^1(Q))^*$ in the sense that it is a bounded linear functional on the space of $1$-forms. The natural duality pairing on $k$-forms is denoted $\langle \cdot, \cdot \rangle_{\Lambda^k} : (\Lambda^k(Q))^* \times \Lambda^k(Q) \to \mathbb{R}$. This is not the pointwise duality of vectors and covectors, but rather duality at the level of the function space. Because $\delta F/\delta \bm{d}^1_* \in \Lambda^1(Q)$ and $\bm{b}^2 \in \Lambda^2(Q)$, it follows that we may rewrite the particle coupling terms as
\begin{equation}
	\nabla_{\bm{u}} \frac{\delta F}{\delta f} \cdot \frac{\delta G}{\delta \bm{D}} \mapsto 
	\bm{\iota}_{\mathsf{d}_{\bm{u}} \frac{\delta F}{\delta f}} \frac{\delta G}{\delta \bm{d}^1_*} = \frac{\delta G}{\delta \bm{d}^1_*} \left( \mathsf{d}_{\bm{u}} \frac{\delta F}{\delta f} \right)
\end{equation}
and
\begin{equation}
	\bm{B} \cdot \nabla_{\bm{u}} \frac{\delta F}{\delta f} \times \nabla_{\bm{u}} \frac{\delta G}{\delta f} \mapsto 
	\bm{\iota}_{ \mathsf{d}_{\bm{u}} \frac{\delta F}{\delta f}} \bm{\iota}_{\mathsf{d}_{\bm{u}} \frac{\delta G}{\delta f}} \bm{b}^2 =
	 \bm{b}^2 \left( \mathsf{d}_{\bm{u}} \frac{\delta F}{\delta f}, \mathsf{d}_{\bm{u}} \frac{\delta G}{\delta f} \right)
\end{equation}
where on the left the functional derivatives are understood with respect to the $L^2$ pairing, and on the right, with respect to the natural pairing via functional evaluation. 

\subsection{The geometric and Hamiltonian structure of Vlasov-Maxwell}
Using the results for the macroscopic Maxwell equations from section \ref{ssec:genpair} and the previous subsection, we find that the geometric Vlasov-Maxwell Hamiltonian structure may be written
\begin{equation}
	\begin{aligned}
		\{F, G\} = 
		&\frac{1}{m} \left\langle f, \left[ \frac{\delta F}{\delta f}, \frac{\delta G}{\delta f} \right] \right\rangle_{\mathfrak{g}^*, \mathfrak{g}} 
		+ \frac{4 \pi q}{m} \left\langle f, \frac{\delta G}{\delta \bm{d}^1_*} \left( \mathsf{d}_{\bm{u}} \frac{\delta F}{\delta f} \right) - \frac{\delta F}{\delta \bm{d}^1_*} \left( \mathsf{d}_{\bm{u}} \frac{\delta G}{\delta f} \right) \right\rangle_{\mathfrak{g}^*, \mathfrak{g}} \\
		+ &\frac{q}{m^2 c} \left\langle f,  \bm{b}^2 \left( \mathsf{d}_{\bm{u}} \frac{\delta F}{\delta f}, \mathsf{d}_{\bm{u}} \frac{\delta G}{\delta f} \right) \right\rangle_{\mathfrak{g}^*, \mathfrak{g}} \\
		- &4 \pi c \left[ \left\langle \frac{\delta F}{\delta \bm{b}^2}, \mathsf{d}_1 \frac{\delta G}{\delta \bm{d}^1_*} \right\rangle_{(\Lambda^2)^*, \Lambda^2} - 
			\left\langle \frac{\delta G}{\delta \bm{b}^2}, \mathsf{d}_1 \frac{\delta F}{\delta \bm{d}^1_*} \right\rangle_{(\Lambda^2)^*, \Lambda^2} \right].
	\end{aligned}
\end{equation}
As desired, this bracket is explicitly metric free. As noted in \cite{eldred_and_bauer}, the Poisson bracket is metric free in general and it should be possible to find an explicitly metric free formulation of the bracket for any Hamiltonian field theory. 

The Hamiltonian is written
\begin{equation}
	H[f, \bm{e}^1, \bm{b}^2] = K[f, \bm{e}^1, \bm{b}^2] - \left\langle \frac{\delta K}{\delta \bm{e}^1}, \bm{e}^1 \right\rangle_{(\Lambda^1)^*, \Lambda^1} + \frac{1}{8 \pi} \left[ \left(\bm{e}^1, \bm{e}^1 \right)_{L^2 \Lambda^1} + \left( \bm{b}^2, \bm{b}^2 \right)_{L^2 \Lambda^2} \right]\,,
\end{equation}
where we define $\bm{d}^1_*$ as in the previous section. We reiterate that we have left the precise notion of duality unspecified for generality, however usually the $L^2$ inner product is used. Moreover, as in \cite{morrison_gauge_free_lifting}, 
\begin{equation}
	K[f, \bm{e}^1, \bm{b}^2] = \left\langle f, \mathcal{K} \right\rangle_{\mathfrak{g}^*, \mathfrak{g}}
\end{equation}
where $\mathcal{K} = \mathcal{K}(\bm{x}, \bm{u}, \bm{e}^1, \bm{b}^2) \in \mathfrak{g}$ for given $\bm{e}^1$ and $\bm{b}^2$. Letting $\overline{H}[f, \bm{d}_*^1, \bm{b}^2] = H[f, \bm{e}^1, \bm{b}^2]$, we find 
\begin{equation}
	\frac{\delta \overline{H}}{\delta \bm{d}^1_*} = \frac{\bm{e}^1}{4 \pi} 
	\quad \text{and} \quad
	\frac{\delta \overline{H}}{\delta \bm{b}^2} = \frac{\bm{h}^2_*}{4 \pi} \,, 
\end{equation}
where $\bm{h}^2_* \in (\Lambda^2(Q))^*$. As shown in \cite{morrison_gauge_free_lifting}, $\delta \overline{H}/ \delta f = \mathcal{K}$. 

\subsection{The weak equations of motion}
Using the expressions for the derivatives of the Hamiltonian, it is possible to derive the equations of motion, viz. 
\begin{align}
		\{F, \overline{H} \} &= 
		\frac{1}{m} \left\langle f, \left[ \frac{\delta F}{\delta f}, \mathcal{K} \right] \right\rangle_{\mathfrak{g}^*, \mathfrak{g}} 
		\!+ \frac{q}{m} \left\langle f, \bm{e}^1 \left( \mathsf{d}_{\bm{u}} \frac{\delta F}{\delta f} \right) 
			- 4 \pi \frac{\delta F}{\delta \bm{d}^1_*} \left( \mathsf{d}_{\bm{u}} \mathcal{K} \right) \right\rangle_{\mathfrak{g}^*, \mathfrak{g}} \\
		&+\frac{q}{m^2 c} \left\langle f,  \bm{b}^2 \left( \mathsf{d}_{\bm{u}} \frac{\delta F}{\delta f}, \mathsf{d}_{\bm{u}} \mathcal{K} \right) \right\rangle_{\mathfrak{g}^*, \mathfrak{g}} \\
		&- c \left[ \left\langle \frac{\delta F}{\delta \bm{b}^2}, \mathsf{d}_1 \bm{e}^1 \right\rangle_{(\Lambda^2)^*, \Lambda^2} - 
			\left\langle \bm{h}^2_*, \mathsf{d}_1 \frac{\delta F}{\delta \bm{d}^1_*} \right\rangle_{(\Lambda^2)^*, \Lambda^2} \right].
\nonumber
\end{align}
Hence, we obtain the weak equations of motion, 
\begin{align}
	\dot{F}[f] &= \frac{1}{m} \left\langle f, \left[ \frac{\delta F}{\delta f}, \mathcal{K} \right] + 
		q \left[ \bm{e}^1 \left( \mathsf{d}_{\bm{u}} \frac{\delta F}{\delta f} \right) 
		+ \frac{1}{m c} \bm{b}^2 \left( \mathsf{d}_{\bm{u}} \frac{\delta F}{\delta f}, \mathsf{d}_{\bm{u}} \mathcal{K} 
		\right) \right] \right\rangle_{\mathfrak{g}^*, \mathfrak{g}},
\\
	\dot{F}[\bm{d}^1_*] &= c \left\langle \bm{h}^2_*, \mathsf{d}_1 \frac{\delta F}{\delta \bm{d}^1_*} \right\rangle_{(\Lambda^2)^*, \Lambda^2}
		- \frac{4 \pi q}{m} \left\langle f, \frac{\delta F}{\delta \bm{d}^1_*} \left( \mathsf{d}_{\bm{u}} \mathcal{K} \right) \right\rangle_{\mathfrak{g}^*, \mathfrak{g}},
\\
	\dot{F}[\bm{b}^2] &= - c \left\langle \frac{\delta F}{\delta \bm{b}^2}, \mathsf{d}_1 \bm{e}^1 \right\rangle_{(\Lambda^2)^*, \Lambda^2}.
\end{align}
These are supplemented with the constitutive relations, 
\begin{equation}
	\bm{d}^1_* = \mathcal{R}_1 \bm{e}^1 - 4 \pi \frac{\delta K}{\delta \bm{e}^1}
	\quad \text{and} \quad
	\bm{h}^2_* = \mathcal{R}_2 \bm{b}^2 + 4 \pi \frac{\delta K}{\delta \bm{b}^2}.
\end{equation}
We think of these constitutive relations as being a part of the Hamiltonian. 

Further simplification is only practical if we prescribe a particular duality pairing in the above formulas. For example, $L^2$ duality reduces the above to what was given in \cite{morrison_gauge_free_lifting}. This weak manner of writing the equations, while inconveniently intricate for certain purposes, has the advantage of explicitly splitting the theory into components which are metric independent (the Poisson bracket) and components which are metric dependent (the Hamiltonian). 

\section{Relativistic Vlasov-Maxwell in field dependent media}
We now consider the behavior of the Vlasov-Maxwell system in electromagnetic field dependent media under Lorentz transformations. The media under consideration are not as general as those considered in \cite{morrison_gauge_free_lifting}, but still accommodate interesting models.

\subsection{Derivation of a Lorentz invariant formulation}
We start by writing out the equations of motion in the more standard language of vector calculus and verifying that they are Lorentz invariant. Let $f = f(\bm{x}, \bm{u})$,  where $\bm{u}$ is the ``reduced" velocity, which is related to the kinematic velocity by
\begin{equation}
	\bm{u} = \frac{\bm{v}/c}{\sqrt{1 - v^2/c^2}} \iff \bm{v} = \frac{ c \bm{u}}{\sqrt{1 + u^2}}.
\end{equation}
We wish to study the following kinetic model:
\begin{equation} \label{eq:relativistic_vlasov-maxwell_1}
	\begin{aligned}
		\frac{\partial f}{\partial t} &+ \frac{c \bm{u}}{\sqrt{1 + u^2}} \cdot \nabla_{\bm{x}} f + \frac{q}{m} \left( \bm{E} + \frac{\bm{u}}{\sqrt{1 + u^2}} \times \bm{B} \right) \cdot \nabla_{\bm{u}} f = 0 \\
		\frac{\partial \bm{D}}{\partial t} &= c \nabla \times \bm{H} - \frac{4 \pi q}{m} \int_Q \frac{c \bm{u}}{\sqrt{1 + u^2}} f \, \mathsf{d}^3 \bm{u} \\
		\frac{\partial \bm{B}}{\partial t} &= - c \nabla \times \bm{E}.
	\end{aligned}
\end{equation}
If we were to let $\bm{D} = \bm{E}/4\pi$ and $\bm{H} = \bm{B}/4\pi$, then these equations reduce to the usual Vlasov-Maxwell equations which are Lorentz invariant (e.g.,  see \cite{ClemmowP.C.1971EoPa}). Hence, we need only consider the assumptions on the constitutive relations for $(\bm{D}, \bm{H})$ which ensure the covariance of Amp{\`e}re's law. 

An idea for building in such covariance for electromagnetic fields in media dates to the early 20th century by Mie, Schwarzschild, and others (cf.\  \cite{sommerfeld3}).  One proceeds by building  a  Lagrangian density out of Lorentz invariant terms,  and then obtains constitutive relations by taking partial derivatives of the Lagrangian density with respect to the  fields $\bm{E}$ and $\bm{B}$.  In the spirit of the polarization and magnetization calculations of \cite{morrison_gauge_free_lifting},  we generalize this procedure by writing the constitutive relations in terms of functional derivatives of the Lagrangian, which allows for the accommodation of more general theories that involve higher order derivative Lagrangians (e.g.,\cite{pegoraro-bulanov}). 

For the present context, we suppose an  electromagnetic Lagrangian of the following form:
\begin{equation}
    L_{EM}[\bm{E}, \bm{B}] = \frac{1}{2} \int_Q \left( | \bm{E} |^2 - | \bm{B} |^2 \right) \mathsf{d}^3 \bm{x}
                        - 4 \pi K_{EM}[\bm{E}, \bm{B}]\,,
\end{equation}
where $K_{EM}$ is an arbitrary functional  the fields.  Then we define
\begin{equation}
	\bm{D} = \frac{\delta L_{EM}}{\delta \bm{E}} = \bm{E} - 4 \pi \frac{\delta K_{EM}}{\delta \bm{E}}
	\quad \text{and} \quad 
	\bm{H} = - \frac{\delta L_{EM}}{\delta \bm{B}} = \bm{B} + 4 \pi \frac{\delta K_{EM}}{\delta \bm{B}}\,,
	\label{genform}
\end{equation}
as in  \cite{morrison_gauge_free_lifting}. 

The Lagrangian transforms as a scalar between inertial reference frames $S$ and $S'$ with relative velocity $\bm{v}$:
\begin{equation}
	L_{EM}[\bm{E}, \bm{B}] = L_{EM}'[\bm{E}', \bm{B}'] = 
	\frac{1}{2} \int_Q \left( | \bm{E}' |^2 - | \bm{B}' |^2 \right) \mathsf{d}^3 \bm{x} - 4 \pi K_{EM}'[\bm{E}', \bm{B}']\,,
\end{equation}
where we let $K_{EM}'[\bm{E}', \bm{B}' ] = K_{EM}[\bm{E}, \bm{B}]$. This may be accomplished by making $K_{EM}$ depend only on the fields through the two Lorentz invariants, $|\bm{E}|^2-|\bm{B}|^2$ and $\bm{E}\cdot \bm{B}$,  so that $K_{EM}$ itself is Lorentz invariant. The Lorentz boosted fields may be written as 
\begin{equation}
	\begin{pmatrix}
		\bm{E}' \\
		\bm{B}'
	\end{pmatrix}
		=
	\begin{pmatrix}
		\gamma \mathbb{I} + (1 - \gamma) \dfrac{\bm{v} \otimes \bm{v}}{v^2} & \gamma \hat{\bm{v}} \\
		- \dfrac{\gamma}{c^2} \hat{\bm{v}} & \gamma \mathbb{I} + (1 - \gamma) \dfrac{\bm{v} \otimes \bm{v}}{v^2} 
	\end{pmatrix}
	\begin{pmatrix}
		\bm{E} \\
		\bm{B}
	\end{pmatrix}
		=: \mathbb{B}(\bm{v})
	\begin{pmatrix}
		\bm{E} \\
		\bm{B}
	\end{pmatrix}\,,
\end{equation}
where the {\it hat} map indicates $\hat{\bm{v}} \bm{B} = \bm{v} \times \bm{B}$. 

Therefore, it follows that 
\begin{align}
	L_{EM}'[\bm{E}', \bm{B}'] &= L_{EM}'[\mathbb{B}(\bm{v})( \bm{E}, \bm{B} )] = L_{EM}[\bm{E}, \bm{B} ] 
	\quad \implies
	\nonumber\\
	 L_{EM}'[\bm{E}', \bm{B}'] &= L_{EM}[\mathbb{B}(-\bm{v})( \bm{E}', \bm{B}' )]
	\,,
\end{align}
since $\mathbb{B}^{-1}(\bm{v}) = \mathbb{B}(-\bm{v})$, and  we find that
\begin{equation}
	\begin{aligned}
		\frac{\delta L_{EM}}{\delta (\bm{E}, \bm{B} )} \cdot (\delta \bm{E}, \delta \bm{B} ) 
			&= \frac{\delta L_{EM}'}{\delta (\bm{E}', \bm{B}'  )} \cdot (\delta \bm{E}', \delta \bm{B}' ) \\
			&= \frac{\delta L_{EM}'}{\delta ( \bm{E}', \bm{B}'  )} \cdot \mathbb{B}(\bm{v}) ( \delta \bm{E}, \delta \bm{B} ) \\
			&= \mathbb{B}^T(\bm{v}) \frac{\delta L_{EM}'}{\delta ( \bm{E}', \bm{B}'  )} ( \delta \bm{E}, \delta \bm{B} )\,,
	\end{aligned}
\end{equation}
which implies
\begin{equation}
	\frac{\delta L_{EM}}{\delta (\bm{E}, \bm{B} )} = \mathbb{B}^T(\bm{v}) \frac{\delta L_{EM}'}{\delta ( \bm{E}', \bm{B}'  )}.
\end{equation}
Therefore, we find that
\begin{equation}
	\begin{pmatrix}
		\bm{D}' \\
		-\bm{H}'
	\end{pmatrix}
		=
	\frac{\delta L_{EM}}{\delta ( \bm{E}', \bm{B}' )} = \mathbb{B}^{-T}(\bm{v}) \frac{\delta L_{EM}}{\delta ( \bm{E}, \bm{B} )}
		=
	\mathbb{B}^{-T}(\bm{v})
	\begin{pmatrix}
		\bm{D} \\
		-\bm{H}
	\end{pmatrix}.
\end{equation}
However, 
\begin{equation}
	\mathbb{B}^{-T}(\bm{v}) 
		=
	\mathbb{B}^{T}(-\bm{v}) 
		=
	\begin{pmatrix}
		\gamma \mathbb{I} + (1 - \gamma) \dfrac{\bm{v} \otimes \bm{v}}{v^2} & -\dfrac{\gamma}{c^2} \hat{\bm{v}} \\
		\gamma \hat{\bm{v}} & \gamma \mathbb{I} + (1 - \gamma) \dfrac{\bm{v} \otimes \bm{v}}{v^2} 
	\end{pmatrix}\,,
\end{equation}
since $\hat{\bm{v}}^T = -\hat{\bm{v}}$. Hence,
\begin{equation}
	\begin{aligned}
		\begin{pmatrix}
			\bm{D}' \\
			-\bm{H}'
		\end{pmatrix}
			&=
		\begin{pmatrix}
			\gamma \mathbb{I} + (1 - \gamma) \dfrac{\bm{v} \otimes \bm{v}}{v^2} & -\dfrac{\gamma}{c^2} \hat{\bm{v}} \\
			\gamma \hat{\bm{v}} & \gamma \mathbb{I} + (1 - \gamma) \dfrac{\bm{v} \otimes \bm{v}}{v^2} 
		\end{pmatrix}
		\begin{pmatrix}
			\bm{D} \\
			-\bm{H}
		\end{pmatrix} \\
			&=
		\begin{pmatrix}
			\gamma(\bm{D} + \dfrac{1}{c^2} \bm{v} \times \bm{H}) + (1 - \gamma) \dfrac{\bm{v} \cdot \bm{D}}{v^2} \bm{v} \\
			\gamma(- \bm{H} + \bm{v} \times \bm{D}) - (1 - \gamma) \dfrac{\bm{v} \cdot \bm{H}}{v^2} \bm{v}
		\end{pmatrix}\,,
	\end{aligned}
\end{equation}
which upon simplification  gives
\begin{equation}
	\begin{aligned}
		\bm{D}' &= \gamma(\bm{D} + \dfrac{1}{c^2} \bm{v} \times \bm{H}) + (1 - \gamma) \dfrac{\bm{v} \cdot \bm{D}}{v^2} \bm{v} \\
		\bm{H}' &= \gamma(\bm{H} - \bm{v} \times \bm{D}) + (1 - \gamma) \dfrac{\bm{v} \cdot \bm{H}}{v^2} \bm{v}.
	\end{aligned}
	\label{results}
\end{equation}
Equations (\ref{results}) describe  precisely the manner in which the macroscopic fields must transform to ensure the Lorentz invariant of Maxwell's equations. 

If one defines the constitutive relation between $(\bm{E},\bm{B})$ and $(\bm{D}, \bm{H})$ via our general  functional derivative form of \eqref{genform} where the  Lagrangian is an arbitrary functional of  Lorentz invariants, the resulting kinetic theory is Lorentz invariant. As many models in nonlinear electrodynamics are prescribed via a Lorentz invariant  Lagrangian, e.g., the Born-Infeld and Euler-Heisenberg models and those of \cite{pegoraro-bulanov}, this framework provides a convenient means of coupling such models to a plasma. 

\subsection{Derivation of the Hamiltonian structure}
As just demonstrated, the electromagnetic Lagrangian as defined in the previous section is Lorentz invariant and therefore provides a convenient starting point to define the Hamiltonian. The Hamiltonian is defined via a Legendre transform of the electromagnetic Lagrangian plus the relativistic kinetic energy:
\begin{equation} \label{eq:relativistic_hamiltonian}
	\begin{aligned}
    		H[f, \bm{E}, \bm{B}] 
		&= \int_{TQ} mc \sqrt{1 + u^2} f(\bm{x}, \bm{u})\,  \mathsf{d}^3 \bm{x}\, \mathsf{d}^3 \bm{u} + \int_Q \frac{\delta L_{EM}}{\delta \bm{E}} \cdot \bm{E}\,  \mathsf{d}^3 \bm{x} 
			- \frac{L_{EM}[\bm{E}, \bm{B}]}{4 \pi} \\
    		&= \int_{TQ} mc \sqrt{1 + u^2} f(\bm{x}, \bm{u})\,  \mathsf{d}^3 \bm{x}\,  \mathsf{d}^3 \bm{u} 
			+ K_{EM}[\bm{E}, \bm{B}] \\
		&\hspace{5em} + \int_Q \bm{D} \cdot \bm{E}\,  \mathsf{d}^3 \bm{x} 
			- \frac{1}{8 \pi} \int_Q \left[ |\bm{E}|^2 - | \bm{B} |^2 \right] \, \mathsf{d}^3 \bm{x}\,,
	\end{aligned}
\end{equation}
since
\begin{equation}
	\bm{D} = \frac{\delta L_{EM}}{\delta \bm{E}} = \bm{E} - 4 \pi \frac{\delta K_{EM}}{\delta \bm{E}}.
\end{equation}
One can clearly see that this reduces to the form of Hamiltonian prescribed in \cite{morrison_gauge_free_lifting}. A similar Hamiltonian for a model with point charges is defined in \cite{IBB73}, and is shown to arise from a variational principle. 

While the electromagnetic Lagrangian is Lorentz invariant, $L'_{EM}[f', \bm{E}', \bm{B}'] = L_{EM}[f,\bm{E},\bm{B}]$, it is not in general the case that $H'[f', \bm{E}', \bm{B}']=H[f,\bm{E},\bm{B}]$. Rather, in a given inertial reference frame, we subordinate the definition of the Hamiltonian to that of the Lagrangian:
\begin{align}
    		H'[f', \bm{E}', \bm{B}'] 
		&= \int_{TQ} mc \sqrt{1 + u'^2} f'(\bm{x}', \bm{u}') \mathsf{d}^3 \bm{x}' \, \mathsf{d}^3 \bm{u}' \\
		&\hspace{5em}+ \int_Q \frac{\delta L_{EM}'}{\delta \bm{E}'} \cdot \bm{E}' \, \mathsf{d}^3 \bm{x}' 
			- \frac{L_{EM}'[\bm{E}', \bm{B}']}{4 \pi}
			\nonumber \\
    		&= \int_{TQ} mc \sqrt{1 + u'^2} f'(\bm{x}', \bm{u}')\,  \mathsf{d}^3 \bm{x}' \mathsf{d}^3 \bm{u}' 
			+ K_{EM}'[\bm{E}', \bm{B}']
			\nonumber \\
		&\hspace{5em} + \int_Q \bm{D}' \cdot \bm{E}' \mathsf{d}^3 \bm{x} 
			- \frac{1}{8 \pi} \int_Q \left[ |\bm{E}'|^2 - | \bm{B}' |^2 \right] \mathsf{d}^3 \bm{x}'.
	\end{align}
The Poisson bracket is defined in the same manner regardless of our choice of reference frame. It is straightforward to show that this Hamiltonian along with the Poisson bracket from \cite{morrison_gauge_free_lifting} yields the equations of motion given in  Eqs.~(\ref{eq:relativistic_vlasov-maxwell_1}).  It should be noted that in the nonrelativistic limit the Hamiltonian theory clearly reduces to that of \cite{morrison_gauge_free_lifting}, since this is immediate from the nonrelativistic limit of the Hamiltonian, which reduces for simple media to the original Vlasov-Maxwell Hamiltonian structure of \cite{morrison-1980_maxwell-vlasov,morrison_poisson_brackets,MARSDEN1982394}.  Similarly, for relativistic simple media, the  theory reduces to that given in  \cite{BIALYNICKIBIRULA1984509}. 
\subsection{Geometric Relativistic Vlasov-Maxwell}
We now briefly summarize the expression of the Lorentz-invariant relativistic Vlasov-Maxwell system discussed in the previous sections in terms of the geometric language developed in this paper. The pointwise kinetic energy is given by
\begin{equation}
	\mathcal{K} = m c \sqrt{1 + u^2 }
	\quad \text{where} \quad
	\bm{u} = \frac{\bm{v}/c}{\sqrt{1 - v^2/c^2}}
\end{equation}
is the reduced velocity and $\bm{v}$ is the kinematic velocity. One can see that
\begin{equation}
	\mathsf{D}_{\bm{u}} \mathcal{K} = \frac{\partial \mathcal{K}}{\partial \bm{u}} \cdot \frac{\partial}{\partial \bm{x}} = \frac{m c \bm{u}}{\sqrt{1 + u^2}} \cdot \frac{\partial}{\partial \bm{x}}
\end{equation}
where the dot-product notation indicates contraction of up and down indices. 

For notational simplicity, we identify the dual space via the $L^2$ inner product. Hence, all functional derivatives in the following will be identified with respect to the $L^2$ inner product. Denote $\bm{e}^1 = \bm{E}(\bm{x}) \cdot \mathsf{d} \bm{x}$, $\bm{b}^2 = \bm{B} \cdot \mathsf{d} \bm{S}$, 
\begin{equation}
	\bm{d}^1 = \frac{\delta L_{EM}}{\delta \bm{e}^1} = \bm{D} \cdot \mathsf{d} \bm{x},
	\quad \text{and} \quad
	\bm{h}^2 = \frac{\delta L_{EM}}{\delta \bm{b}^2} = \bm{H} \cdot \mathsf{d} \bm{S}\,,
\end{equation}
where the electromagnetic Lagrangian $L_{EM}[\bm{e}^1, \bm{b}^2]$ is assumed to be Lorentz invariant. We find
\begin{equation}
	\begin{aligned}
		\{F, H\} = 
		&\left( f, \frac{\partial}{\partial \bm{x}} \frac{\delta F}{\delta f} \cdot  \frac{ c \bm{u}}{\sqrt{1 + u^2}} \right)_{L^2(T^*Q)}
		+ \frac{q}{m c} \left( f,  \bm{B} \cdot \left( \frac{\partial}{\partial \bm{u}} \frac{\delta F}{\delta f} \times  \frac{c \bm{u}}{\sqrt{1 + u^2}} \right) \right)_{L^2(T^*Q)} \\
		&+ \frac{q}{m} \left( f, \bm{E} \cdot \frac{\partial}{\partial \bm{u}} \frac{\delta F}{\delta f}
			- 4 \pi \frac{\delta F}{\delta \bm{D}} \cdot \frac{ m c \bm{u}}{\sqrt{1 + u^2}} \right)_{L^2(T^*Q)} \\
		- &c \left[ \left( \frac{\delta F}{\delta \bm{B}}, \nabla \times \bm{E} \right)_{L^2(Q)} - 
			\left( \bm{H}, \nabla \times \frac{\delta F}{\delta \bm{D}} \right)_{L^2(Q)} \right].
	\end{aligned}
\end{equation}
For arbitrary (time-independent) test functions $g(\bm{x}, \bm{u})$, $\bm{\psi}(\bm{x})$, and $\bm{\phi}(\bm{x})$, let
\begin{equation}
	F[f, \bm{e}^1, \bm{b}^2] = \int_{T^*Q} g f\,  \mathsf{d}^3 \bm{x} \mathsf{d}^3 \bm{u} 
						+ \int_Q \bm{\psi} \cdot \bm{E}\,  \mathsf{d}^3 \bm{x} 
						+ \int_Q \bm{\phi} \cdot \bm{B}\,  \mathsf{d}^3 \bm{x}.
\end{equation}
We may extract each of the three dynamical equations by setting two of the three test functions identically equal to zero. Doing so, we obtain the Vlasov equation,
\begin{equation}
	\left( g, \partial_t f \right)_{L^2(T^*Q)} - \left( f, \frac{c \bm{u}}{\sqrt{1 + u^2}} \cdot \frac{\partial g}{\partial \bm{x}} 
	+ \frac{q}{m} \left[ \bm{E} + \frac{\bm{u}}{\sqrt{1 + u^2}} \times \bm{B} \right] \cdot \frac{\partial g}{\partial \bm{u}} \right)_{L^2(T^*Q)} = 0,
\end{equation}
Faraday's law, $\partial_t \bm{B} = - \nabla \times \bm{E}$, and Amp{\`e}re's law,
\begin{equation}
	\left( \bm{\psi}, \partial_t \bm{D} \right)_{L^2(Q)} = c \left( \bm{H}, \nabla \times \bm{\psi} \right)_{L^2(Q)} - 4 \pi q \left( f, \frac{\bm{\psi} \cdot c \bm{u}}{\sqrt{ 1 + u^2}} \right)_{L^2(T^* Q)}.
\end{equation}
Notice, Faraday's law is expressed strongly whereas the Vlasov equation and Amp{\`e}re's law are expressed weakly. With homogeneous boundary conditions, one may recover the strong equations via integration by parts. It is however useful to have the equations in weak form as this frequently provides a starting place for numerical methods. 

It is clear that $\nabla \cdot \bm{B} = 0$ is exactly conserved by the flow. If we take the functional
\begin{equation}
	C_D[f, \bm{D}] = \int_Q \nabla \eta \cdot \bm{D} + \eta \left( 4 \pi q \int_{T^*_{\bm{x}}Q} f \mathsf{d}^3 \bm{u} \right) \mathsf{d}^3 \bm{x}
\end{equation}
where $\eta = \eta(\bm{x})$ is arbitrary, then we find
\begin{equation}
	\partial_t C_D = \int_{\partial Q} \eta \bm{J} \cdot \mathsf{d} \bm{S} 
	\quad \text{where} \quad
	\bm{J}(\bm{x}) = 4 \pi q \int_{T^*_{\bm{x}} S} \frac{c \bm{u}}{\sqrt{1 + u^2}} f \,  \mathsf{d}^3 \bm{u}.
\end{equation}
Conservation of $C_D$ represents charge conservation in a weak form. Note, only we showed here that this functional Poisson commutes with the Hamiltonian, but on an infinite domain or on a compact Riemannian manifold (without boundary), this is in fact a Casimir invariant \cite{Chandre_2013, Morrison+1987+1115+1123, PhysRevA.40.3898}. Further commentary on boundary conditions is impeded because the appropriate boundary conditions for the distribution function which yield a valid Poisson bracket on a manifold with a boundary are yet unknown. 

\subsection{On the Lorentz invariance of more general media} \label{sec:more_general_media}
Following the approach taken in \cite{ClemmowP.C.1971EoPa}, we now investigate the feasibility of a covariant formulation in more general media. The Vlasov equation obtained by the fully general media prescribed in \cite{morrison_gauge_free_lifting} is
\begin{equation}
	\frac{\partial f}{\partial t} + [f, \mathcal{K}]
		+ \left( \bm{E} + \frac{1}{c} \nabla_{\bm{u}} \mathcal{K} \times \bm{B} \right) \cdot \nabla_{\bm{u}} f = 0
\end{equation}
where $\mathcal{K} = \mathcal{K}(\bm{x}, \bm{u}, \bm{E}, \bm{B})$. For convenience, we have set $m=q=1$. For the sake of simplicity, suppose $\mathcal{K}$ does not depend on $\bm{x}$. Then we find
\begin{equation}
	\frac{\partial f}{\partial t} + \nabla_{\bm{u}} \mathcal{K} \cdot \nabla_{\bm{x}} f
		+ \left( \bm{E} +  \frac{1}{c} \nabla_{\bm{u}} \mathcal{K} \times \bm{B} \right) \cdot \nabla_{\bm{u}} f = 0.
\end{equation}
Let 
\begin{equation}
	\bm{A} = \bm{E} + \frac{1}{c} \nabla_{\bm{u}} \mathcal{K} \times \bm{B}.
\end{equation}
To begin, multiply the entire equation by $\sqrt{1 + u^2}$:
\begin{equation}
	\sqrt{1 + u^2} \frac{\partial f}{\partial t} + \sqrt{1 + u^2} \nabla_{\bm{u}} \mathcal{K} \cdot \nabla_{\bm{x}} f
		+ \sqrt{1 + u^2} \bm{A} \cdot \nabla_{\bm{u}} f = 0.
\end{equation}
As shown in \cite{VANKAMPEN1969244}, $f(\bm{x}, \bm{u})$ is a Lorentz invariant: $f(\bm{x}, \bm{u}) = f'(\bm{x}', \bm{u}')$. Hence, $(\partial_t f, c \nabla_{\bm{x}} f)$ is a covariant $4$-vector. Therefore, in order for the first two terms in the Vlasov equation to transform covariantly, we would need that
\begin{equation}
	\mathfrak{u} = \left( \sqrt{ 1 + u^2 }, \frac{\sqrt{1 + u^2}}{c} \nabla_{\bm{u}} \mathcal{K} \right)
\end{equation}
be a $4$-vector. 

If $F$ is the Faraday tensor, then
\begin{equation}
	F \mathfrak{u} = \left( \bm{A} \cdot \frac{\sqrt{1 + u^2}}{c} \nabla_{\bm{u}} \mathcal{K}, \sqrt{1 + u^2} \bm{A} \right).
\end{equation}
This is a $4$-vector if and only if $\mathfrak{u}$ is a $4$-vector. Beyond $\mathfrak{u}$ being a $4$-vector however, for the kinetic equation to be Lorentz invariant, we would need to show that
\begin{equation} \label{eq:force_lorentz_trans}
	\sqrt{1 + u'^2} \bm{A}' \cdot \nabla_{\bm{u}'} f' = \sqrt{1 + u^2} \bm{A} \cdot \nabla_{\bm{u}} f. 
\end{equation}
Even with the assumption that $\mathfrak{u}$ is a $4$-vector, which is hardly guaranteed, we find that, if $V$ is a boost in the first coordinate direction, 
\begin{equation}
	\sqrt{1 + u'^2} A_1' = \frac{\sqrt{1 + u^2} A_1 - (V/c) \bm{A} \cdot \frac{\sqrt{1 + u^2}}{c} \nabla_{\bm{u}} \mathcal{K} }{\sqrt{1 - V^2/c^2}},
\end{equation}
\begin{equation}
	\sqrt{1 + u'^2} A_i' = \sqrt{1 + u^2} A_i, \quad i = 2,3.
\end{equation}
Moreover,
\begin{equation}
	\frac{\partial f}{\partial u_1'} = \sqrt{ \frac{1 + u^2}{1 + u'^2} } \frac{\partial f}{\partial u_1},
\end{equation}
\begin{equation}
	\frac{\partial f}{\partial u'_i} =  \frac{V/c}{\sqrt{1 - V^2/c^2}}\frac{u_i'}{\sqrt{1 + u'^2}} \frac{\partial f}{\partial u_1} + \frac{\partial f}{\partial u_i}, \quad i = 2,3.
\end{equation}
One can see that the cancelation allowing us to obtain \eqref{eq:force_lorentz_trans} only   if
\begin{equation}
	\frac{\sqrt{1 + u^2}}{c} \nabla_{\bm{u}} \mathcal{K} = \bm{u} \iff \mathcal{K} = c \sqrt{1 + u^2} + C
\end{equation}
where $C$ is a constant of integration. If we relax the requirement that $\mathcal{K}$ not depend on $\bm{x}$, we find that $C$ might depend on $\bm{x}$, $\bm{B}$, and $\bm{E}$. However, the prior arguments remain valid as relaxing the constraint on $\mathcal{K}$ simply adds additional terms to the Vlasov equation (which would likewise would need to transform covariantly). Hence, we conclude that any spatial or field dependence in $\mathcal{K}$ must entirely decouple from those terms with velocity dependence and that the velocity dependence may only appear in the standard form of the relativistic kinetic energy. 

Therefore, in order to obtain a Lorentz invariant kinetic theory from the formalism prescribed in \cite{morrison_gauge_free_lifting}, it follows that it is necessary (but not sufficient) that the energy functional $K$ split as follows:
\begin{equation}
	K[f, \bm{E}, \bm{B}] 
		= \int_{T^*Q} \sqrt{1 + u^2} f \,  \mathsf{d}^3 \bm{x} \mathsf{d}^3 \bm{u} +
		\int_{T^*Q} \mathcal{K}(\bm{x}, \bm{E}, \bm{B}) f \mathsf{d}^3 \bm{x} \mathsf{d}^3\bm{u} 
		+ K_{EM}[\bm{E},\bm{B}]
\end{equation}
which, while slightly generalizing the $K$ functional allowed in equation (\ref{eq:relativistic_hamiltonian}), also places a substantial limitation on the admissible polarizations and magnetizations allowed in a Lorentz invariant kinetic theory. In order for this model to yield a Lorentz invariant theory, $\mathcal{K} = \mathcal{K}(\bm{x}, \bm{E}, \bm{B})$ would need to be such that 
\begin{equation}
	\left( \partial_{\bm{x}} \mathcal{K} + \partial_{\bm{B}} \mathcal{K} \cdot \nabla_{\bm{x}} \bm{B} + \partial_{\bm{E}} \mathcal{K} \cdot \nabla_{\bm{x}} \bm{E} \right) \cdot \nabla_{\bm{u}} f
\end{equation}
remains invariant. Whether there exist such functionals $\mathcal{K}$ remains unclear from this analysis and is beyond the scope of this paper.

\section{Conclusion}
The objectives of this paper were twofold: (1) to express the models from \cite{morrison_gauge_free_lifting} in a geometric language, and (2) to study the conditions for Lorentz invariance in such models. The care taken herein to understand the geometric character of the equations is not done for its own sake, but is primarily accomplished to provide a foundation for future work in structure preserving discretizations of the system. In general, structure preserving discretizations are facilitated by consideration of the geometric structure of the dynamical system \cite{GEMPIC, morrison_structure_preserving_algorithms}. 

The three formulations of Maxwell's equations neatly demonstrate the connection between geometry and discretization. The first formulation based on the double de Rham complex has the advantage of explicitly separating the metric-dependent and independent structures, and transparently represents the geometric character of the equations. However, the formulation involves the explicit use of two distinct duality structures. This formulation most naturally would lead to a numerical strategy based on an explicit discrete Hodge star operator \cite{hiptmair_discrete_hodge_star, kapidani_2022}. The second formulation is based on the $L^2$ inner product. This formulation obfuscates the metric free character of the Poisson bracket, but yields a formulation amenable to methods from finite element exterior calculus \cite{ArnoldDouglasN2010Feec}. The final formulation utilizes an abstract notion of duality and subsumes the previous two. This formulation emphasizes the metric free nature of the Poisson bracket as in the former double de Rham complex formulation while also being a convenient framework for rigorous functional analytic study like the later formulation based on $L^2$ duality. Hence, this final formulation might more easily facilitate the design of structure preserving discretizations.

The second half of the paper considers the full Vlasov-Maxwell system in general media. We first provided a statement of the Poisson bracket in a geometric language utilizing the results from the first half of the paper. The bracket is explicitly metric free and its structure argues for the naturalness of expressing the Vlasov equation and Amp{\`e}re's law in weak form. Finally, we consider the conditions for Lorentz invariance of the kinetic models given in \cite{morrison_gauge_free_lifting}. We find that, while various models from nonlinear electrodynamics based on Lorentz invariant Lagrangians (e.g. Born-Infeld and Euler-Heisenberg electrodynamics) may be coupled to a kinetic model to yield a Lorentz covariant theory, the class of admissible polarizations and magnetizations induced by the plasma itself are somewhat restricted as the Hamiltonian must split in the particular manner described in section \ref{sec:more_general_media}. While this excludes a large number of kinetic models from being Lorentz invariant, a perfectly general prescription remains elusive. 

\section{Acknowledgements}
We gratefully acknowledge the support of U.S. Dept. of Energy Contract \# DE-FG05-80ET-53088, NSF Graduate Research Fellowship \# DGE-1610403, and the Humboldt foundation. PJM would like to acknowledge helpful  conversations with Francesco  Pegoraro. 


\newpage
\addcontentsline{toc}{section}{References}
\bibliographystyle{plain} 
\bibliography{geometric_vlasov.bib} %

\appendix
\section{Notation and mathematical context}
\label{appendix:notation}
\setcounter{equation}{0}

Perhaps the most natural approach to understand the geometric character of the macroscopic Maxwell equations is through the use of exterior calculus. In particular, a form of exterior calculus which distinguishes between twisted and straight differential forms \cite{frankel_2011}. Let $\{ (\Lambda^k, \mathsf{d}_k) \}_{k=0}^n$ be the vector spaces of differential forms on a manifold of dimension $n$.  Here $\Lambda^k$ denotes the set of $k$-forms and $\mathsf{d}_k$ the exterior derivative that takes a $k$-form to a $k+1$-form. We may define a second complex, $\{ (\tilde{\Lambda}^k, \tilde{\mathsf{d}}_k) \}_{k=0}^n$, called the complex of twisted differential forms. This dual complex differs from the first in that twisted forms change sign under orientation changing transformations. The two complexes are related to each other through the Hodge star operator, $\star\colon \Lambda^k \to \tilde{\Lambda}^{n-k}$.  Diagrammatically, this may be expressed as follows:
\begin{equation}
	\begin{tikzcd}
		\cdots \arrow{r} & \Lambda^k \arrow{r}{ \mathsf{d}_k } \arrow{d}{\star} & \Lambda^{k+1} \arrow{r} \arrow{d}{\star} & \cdots \\
		\cdots & \tilde{\Lambda}^{n-k} \arrow{u}  \arrow{l} & \tilde{\Lambda}^{n-(k+1)} \arrow{l}{ \tilde{\mathsf{d}}_{n-(k+1)}} \arrow{u} & \cdots \arrow{l} 
	\end{tikzcd}
\end{equation}
One well-known twisted form is the volume form, $\vol^n$. 

We may define two distinct notions of duality on the double de Rham complex. First, we have the standard $L^2$ inner product, $( \cdot, \cdot )\colon \Lambda^k \times \Lambda^k \to \mathbb{R}$, which is defined
\begin{equation}
	( \omega^k, \eta^k) = \int_Q g^{-k}( \omega^k, \eta^k)\, \vol^n\,,
	\label{L2}
\end{equation}
where $g^{-k}$ is the pointwise inner product on $k$-forms. This pointwise inner product is first defined on $k$-forms that are decomposable into $k$-fold wedge products and then extended to all $k$-forms by linearity. The pointwise inner product when applied to decomposable $k$-forms is the Gram determinant of the inner products of the component $1$-forms:
\begin{equation}
	g^{-k}( \omega_1 \wedge \hdots \wedge \omega_k, \eta_1 \wedge \hdots \wedge \eta_k ) = 
	\det
	\begin{pmatrix}
		( \omega_1, \eta_1 ) & ( \omega_1, \eta_2 ) & \hdots & \\
		(\omega_2, \eta_1) & \ddots & \ddots & \vdots \\
		\vdots & \ddots & \ddots & (\omega_{k-1}, \eta_k) \\
		& \hdots & (\omega_k, \eta_{k-1}) & (\omega_k, \eta_k) 
	\end{pmatrix}.
\end{equation}
Finally, the pointwise inner product of $1$-forms is computed simply via contraction of up and down indicies: 
\begin{equation}
	g^{-1} \left( \omega, \eta \right) = g^{-1} \left( \sum_i \omega_i \mathsf{d} x_i, \sum_j \omega_j \mathsf{d} x_j \right) = \sum_{ij} \omega_i g^{ij} \eta_j \,.
\end{equation}
The second notion of duality is Poincar{\'e} duality, $\langle \cdot, \cdot \rangle\colon \Lambda^k \times \tilde{\Lambda}^{n-k} \to \mathbb{R}$, which is defined
\begin{equation}
	\left\langle \omega^k, \tilde{\eta}^{n-k} \right\rangle_{k,n-k} = \int_Q \omega^k \wedge \tilde{\eta}^{n-k}.
\end{equation}
The $L^2$ inner product, because of its dependence on the Riemannian metric and volume form, is a metric dependent quantity. On the other hand, the Poincar{\'e} duality pairing, built from the wedge product structure alone, is purely topological. Moreover, as both duality pairings are expressed as an integral of a twisted $n$-form, they are independent of the orientation of the coordinate system. The Hodge star operator $\star\colon \Lambda^k \to \tilde{\Lambda}^{n-k}$ is defined such that
\begin{equation}
	\left( \omega^k, \eta^k \right) = \left\langle \omega^k, \star \eta^k \right\rangle_{k,n-k}.
\end{equation}
Note that the Hodge star is not a single operator, but rather a family of operators, one for each $k$. A more precise notation might be $\star_{n-k,k}\colon \Lambda^k \to \tilde{\Lambda}^{n-k}$, however we generally opt for the more concise notation. 

In order to translate vector calculus expressions into the language of differential geometry, it is necessary to invoke the index lowering or flat operator: $( \cdot )^\flat \colon \mathfrak{X} \to \Lambda^1$ defined by
\begin{equation}
	u^1 = g( \cdot, U) = g_{ij} U^j \mathsf{d} x^i := U^\flat\,,
\end{equation}
where $g$ is the metric.  Here we use the superscript $1$ on $u^1$ to indicate that this quantity is a 1-form. The inverse of this operation is the index raising or sharp operator: $( \cdot )^\sharp \colon  \Lambda^1 \to \mathfrak{X}$. We may likewise define an isomorphism between vector fields and twisted $(n-1)$-forms. We define $\textbf{i}_{(\cdot)} \vol^n \colon \mathfrak{X} \to \tilde{\Lambda}^{n-1}$ by
\begin{equation}
	\tilde{u}^{n-1} = \textbf{i}_U \vol^n = \sum_{i} U^i \sqrt{\det(g)} \mathsf{d} x^1 \wedge \hdots \wedge \widehat{\mathsf{d} x^i} \wedge \hdots \wedge \mathsf{d} x^n\,,
	\label{volume}
\end{equation}
where the hat symbol means omission of ``$\mathsf{d} x^i$" from the wedge product and $\textbf{i}_{U} \alpha$ is the interior product of $\alpha$ on  $U$. It is possible to show that $\textbf{i}_U \vol^n = \star U^\flat$. Hence, the inverse operation is given by $U = \left( \star \tilde{u}^{n-1} \right)^\sharp$. It is worth noting that if $U$ is a pseudovector (i.e. a vector which changes sign under orientation reversing transformations), then $\tilde{u}^1 = U^\flat$ is twisted while $u^{n-1} = \textbf{i}_U \vol^n$ is straight. This consideration is important in the case of Maxwell's equations since $\bm{B}$ and $\bm{H}$ are pseudovectors. 

Finally, we note the correspondence of the differential operators from vector calculus with exterior derivatives. If $f\colon Q \to \mathbb{R}$ is a scalar field on a Riemannian manifold, its gradient and exterior derivative are related to each other via
\begin{equation}
	\mathsf{d}_0 f = (\nabla f)^\flat \iff \nabla f = ( \mathsf{d}_0 f )^\sharp.
\end{equation}
Let $U$ be a vector field and $u^1 = U^\flat$. Then the curl of $U$ is defined by
\begin{equation} \label{curl_differential_forms}
	\mathsf{d}_1 u^1 = \normalfont\textbf{i}_{\nabla \times U} \vol^3 = \star ( \nabla \times U)^\flat
	\iff
	\nabla \times U = (\star ( \mathsf{d}_1 u^1 ))^\sharp.
\end{equation}
Letting $\tilde{u}^{2} = \textbf{i}_U \vol^3$, then the divergence is defined to be
\begin{equation}
	\tilde{\mathsf{d}}_2 \tilde{u}^2 
	= \tilde{\mathsf{d}}_2 \textbf{i}_U \vol^3
	= (\nabla \cdot U) \vol^3
	= \star (\nabla \cdot U)
	\iff
	\nabla \cdot U = \star \tilde{\mathsf{d}}_2 \tilde{u}^2.
\end{equation}

As Hamiltonian field theories are formulated via the calculus of variations, it is necessary to briefly consider the calculus of variations with respect to differential forms. Let $V^k$ denote a Hilbert space of differential $k$-forms (and similarly define $\tilde{V}^k$). For example, we might let
\begin{equation}
	V^k = L^2 \Lambda^k(\Omega) = \{ \omega^k \in \Lambda^k \colon \| \omega^k \|_{L^2}^2 = (\omega^k, \omega^k)_{L^2} < \infty \},
\end{equation}
where (for emphasis) only here we added the subscript $L^2$ to the inner product of  \eqref{L2}, or, if we want the exterior derivative to be a bounded operator, 
\begin{equation}
	V^k = H^1 \Lambda^k(\Omega) = \{ \omega^k \in L^2 \Lambda^k(\Omega) \colon \| \mathsf{d}_k \omega^k \|_{L^2} < \infty \}.
\end{equation}
We shall not worry about functional analytic rigor here nor what particular Hilbert space we mean by $V^k$. Rather, we concern ourselves only with the formal correctness of our expressions. Consider a functional  $K\colon V^k \to \mathbb{R}$. We may define a Fr{\'e}chet derivative of this functional in the usual manner
\begin{equation}
	\big| K [\omega + \eta] - K[\omega] - D K[\omega] \eta \big| = O( \| \eta \|)\,.
\end{equation}
Note that $DK[\omega] \in (V^k)^*$, the dual space to the $k$-forms. The manner in which we express the Hamiltonian structure of the macroscopic Maxwell equations is greatly dependent on how we express the dual space. Throughout this paper, we express duality pairings via
\begin{equation}
	DK[\omega] \eta = \left( \frac{\delta K}{\delta \omega}, \eta \right)_{L^2}
	                           = \left\langle \frac{\tilde{\delta} K}{\delta \omega}, \eta \right\rangle_{\tilde{V}^{n-k}, V^k}
			          = \left\langle \frac{\delta K}{\delta \omega}, \eta \right\rangle_{(V^k)^*, V^k}\,,
\end{equation}
where the first is simply $L^2$ duality, the second utilizes the wedge product duality between twisted and straight forms, and the third is the abstract duality pairing via functional evaluation. We notationally distinguish the second variety of functional derivative and call it a ``twisted functional derivative." Notice, the twisted functional derivative of a straight form is a twisted form \cite{eldred_and_bauer}. It is related to the functional derivative identified with $L^2$ duality by the Hodge star operator:
\begin{equation}
	\frac{\delta K}{\delta \omega} = \star \frac{\tilde{\delta} K}{\delta \omega}\,.
\end{equation}
We do not notationally distinguish the functional derivative in the context of abstract duality from the $L^2$ functional derivative because the intended meaning should be clear from context. 

\section{Derivatives of the macroscopic Maxwell Hamiltonian} \label{appendix:deriv-of-ham}
\setcounter{equation}{0}

In order to obtain the equations of motion, we need to take derivatives of the Hamiltonian with respect to $(\bm{D}, \bm{B})$. As the details of this procedure are omitted in \cite{morrison_gauge_free_lifting}, it is useful to show the full calculation here since we will perform the calculation again in the language of exterior calculus in  Sec.~\ref{geoMEs}. Recall,
\begin{equation}
	H[\bm{E},\bm{B}] = K - \int_\Omega \bm{E} \cdot \frac{\delta K}{\delta \bm{E}}\,  \mathsf{d}^3 \bm{x}  + \frac{1}{8 \pi} \int_M ( \bm{E} \cdot \bm{E} + \bm{B} \cdot \bm{B} )\,  \mathsf{d}^3 \bm{x} \,,
\end{equation}
\begin{equation}
	\bm{D} = \bm{E} - 4 \pi \frac{\delta K}{\delta \bm{E}}
	\quad \text{and} \quad
	\bm{H} = \bm{B} + 4 \pi \frac{\delta K}{\delta \bm{B}}\,.
\end{equation}
\begin{lemma}
If we think of $\bm{E}$ as an implicit function of $(\bm{D}, \bm{B})$, then
\begin{equation}
	\frac{\delta \bm{E}}{\delta \bm{D}} = \left( I - 4 \pi \frac{\delta^2 K}{\delta \bm{E} \delta \bm{E}} \right)^{-1}
	\quad \text{and} \qquad
	\frac{\delta \bm{E}}{\delta \bm{B}} = 4 \pi \left( I - 4 \pi \frac{\delta^2 K}{\delta \bm{E} \delta \bm{E}}  \right)^{-1} \frac{\delta^2 K}{\delta \bm{B} \delta \bm{E}}.
\end{equation}
\end{lemma}
\noindent \textit{Proof:} Let $\Phi[ \bm{E}, \bm{B}] = (\bm{D}, \bm{B})$. That is,
\begin{equation*}
	\Phi[ \bm{E}, \bm{B}] = \left( \bm{E} - 4 \pi \frac{\delta K}{\delta \bm{E}}, \bm{B} \right).
\end{equation*}
We assume that $K$ is such that $\Phi$ is a diffeomorphism; hence, we also have $\Phi^{-1}[ \bm{D}, \bm{B}] = (\bm{E}, \bm{B})$.  Upon variation we obtain
\begin{equation*}
	\begin{aligned}
    		D \Phi [\bm{E},\bm{B}](\delta \bm{E}, \delta \bm{B}) 
    		&= 
		\begin{pmatrix}
			D_1 \Phi_1[\bm{E},\bm{B}] & D_2 \Phi_1[\bm{E},\bm{B}] \\
			0 & 1
		\end{pmatrix}
		\begin{pmatrix}
			\delta \bm{E} \\
			\delta \bm{B}
		\end{pmatrix} \\
		&=
		\begin{pmatrix}
			I - 4 \pi \frac{\delta^2 K}{\delta \bm{E} \delta \bm{E}} & - 4 \pi \frac{\delta^2 K}{\delta \bm{B} \delta \bm{E}} \\
			0 & 1
		\end{pmatrix}
		\begin{pmatrix}
			\delta \bm{E} \\
			\delta \bm{B}
		\end{pmatrix}.
	\end{aligned}
\end{equation*}
Hence, it follows that
\begin{equation*}
  	D \Phi^{-1}[\bm{D},\bm{B}](\delta \bm{D}, \delta \bm{B}) 
	= 
	\begin{pmatrix}
		D_1 \Phi_1[\bm{E},\bm{B}]^{-1} & - D_1 \Phi_1[\bm{E},\bm{B}]^{-1} D_2 \Phi_1[\bm{E},\bm{B}] \\
		0 & 1
	\end{pmatrix}
	\begin{pmatrix}
		\delta \bm{D} \\
		\delta \bm{B}
	\end{pmatrix}.
\end{equation*}
Computing the entries of this matrix, we find
\begin{equation*}
 	\frac{\delta \bm{E}}{\delta \bm{D}} = \left( I - 4 \pi \frac{\delta^2 K}{\delta \bm{E} \delta \bm{E}}  \right)^{-1}
	\quad \text{and} \qquad
	\frac{\delta \bm{E}}{\delta \bm{B}} = 4 \pi \left( I - 4 \pi \frac{\delta^2 K}{\delta \bm{E} \delta \bm{E}}  \right)^{-1} \frac{\delta^2 K}{\delta \bm{B} \delta \bm{E}}\,.
\end{equation*}
\qed

\begin{prop}
Let $\overline{H}[\bm{D}, \bm{B}] = H[\bm{E}, \bm{B}]$. Then
\begin{equation} \label{vec_ham_derivs}
	\frac{\delta \overline{H}}{\delta \bm{D}} = \frac{\bm{E}}{4 \pi} \quad \text{and} \quad \frac{\delta \overline{H}}{\delta \bm{B}} = \frac{\bm{H}}{4 \pi}.
\end{equation}
\end{prop}
\noindent \textit{Proof:} Taking derivatives of $H$ with respect to $(\bm{E}, \bm{B})$, we find
\begin{equation*}
	\frac{\delta H}{\delta \bm{E} } = \left( I - 4 \pi \frac{\delta K}{\delta \bm{E} \delta \bm{E}} \right) \frac{\bm{E}}{4 \pi} \quad \text{and} \quad
	\frac{\delta H}{\delta \bm{B} } = \frac{\delta K}{\delta \bm{B}} - \left( \frac{\delta K}{\delta \bm{B} \delta \bm{E}} \right)^* \bm{E} + \frac{\bm{B}}{4 \pi}.
\end{equation*}
The chain rule implies
\begin{equation*}
	\frac{\delta \overline{H}}{\delta \bm{D}} = \left( \frac{\delta \bm{E}}{\delta \bm{D}} \right)^* \frac{\delta H}{\delta \bm{E}} = \left( \frac{\delta \bm{E}}{\delta \bm{D}} \right)^* \left( I - 4 \pi \frac{\delta K}{\delta \bm{E} \delta \bm{E}} \right) \frac{\bm{E}}{4 \pi} = \frac{\bm{E}}{4 \pi}\,.
\end{equation*}
Likewise, 
\begin{align*}
	\frac{\delta \overline{H}}{\delta \bm{B}} 
	&= \frac{\delta H}{\delta \bm{B}} + \left( \frac{\delta \bm{E}}{\delta \bm{B}} \right)^* \frac{\delta H}{\delta \bm{E}} \\
	&= \frac{\delta K}{\delta \bm{B}} - \left( \frac{\delta^2 K}{\delta \bm{B} \delta \bm{E}} \right)^* \bm{E} + \frac{\bm{B}}{4 \pi} 
	+ \left( \frac{\delta \bm{E}}{\delta \bm{B}} \right)^* \left( I - 4 \pi \frac{\delta K}{\delta \bm{E} \delta \bm{E}} \right) \frac{\bm{E}}{4 \pi} \\
	&= \frac{\bm{B}}{4 \pi} + \frac{\delta K}{\delta \bm{B}} = \frac{\bm{H}}{4 \pi}\,.
\end{align*}
\qed

\end{document}